# Image Handling and Processing for Efficient Parking Space Detection and Navigation Aid

Chetan Sai Tutika, Charan Vallapaneni, Karthikeyan B
School of Electronics Engineering
VIT University, India

Email: tchetan.sai2014@vit.ac.in, charanvallapaneni@gmail.com, bkarthikeyan @vit.ac.in

*Abstract*— **This paper aims to develop a robust and flexible algorithm for vacant parking space detections using image processing capabilities of OpenCV. It removes the need for independent sensors to detect a car and instead, uses real-time images derived from various sources and servers to consider a group of slots together. This greatly decreases the expenses required to design an efficient parking system and increases the flexibility of the operation. This method includes the use of a portable processing system with recognition algorithm and has the option of extracting and importing images to the specified servers. The results can be viewed on a custom website with the option to reserve the particular empty slots and GPS navigations to the selected slots.**

## I. INTRODUCTION

Traffic arising from automobiles searching for vacant parking spaces is prominent in populated urban areas. Real-time parking occupation data provides critical input for a parking management system, which are usually acquired by on-site sensors. However, insertion and sustaining of these sensors can be cost inducing. Besides, collaborative mapping and routing optimization for parking with low market penetration is a significant problem for automated vehicles [1]. To solve the parking organization issue, different techniques have been developed and research was conducted to develop efficient parking systems. While some parking systems are deployed as stand-alone technologies, in other situations, multiple techniques are combined to achieve the given task. These technologies include Digital Image processing systems, embedded systems and Internet of Things[2].

The use of IOT with the systems developed, increases the flexibility of the work and gives easy access to the end user. The Internet of Things is a system of interconnected computing devices, mechanical and digital machines, objects, animals or people. The capability to transfer information from one system to other system without the needs of any human resources[3]. Internet of Things (IoT) is the expression first coined by Kevin Ashton executive director of the Auto-ID Center. Today the market scope for IoT is reaching new levels as estimated by 2011 Gartnerâ report [4]. While IOT is highly feasible and flexible the efficiency of such an IoT based system heavily depends on the algorithms which correlate and analyze the collected data through IoT Image data and quality[5].

In this paper, a robust algorithm is designed taking into consideration the joined capabilities of Gaussian blurring, Truncate thresholding, Canny edge detection, contour Detection combined with the capabilities of IOT to give user efficient results and access to data. Image recognition/classification in a brad sense is an extremely diverse area which evidently cannot be determined by a single, optimal method. By use of Gaussian Blur the figure is stripped of high frequency and sharp contrast for smooth analysis[6]. This augments the texture detection of the set image while removing noise added during acquisition the image. Textures from images will give a substantial deal of data about the images[7].

The edge of the image is the extreme at which the gray value of the neighboring pixel at the sudden point of the signal changes drastically. Edge detection is a crucial part of computer vision and image analysis method of contour detection in image processing[8][9]. Edge contours includes rich image information which constitute a characteristic set containing entities different from other image elements. Presently, many applications depend on accurate tracking of target contours obtained through many approaches[10]. It is one of the fundamental and still open image processing tasks due to the complexity of analysis of images of different types with a large number of considered classes of objects[11].

The parking monitoring system proposed is low consumption, easy to implement, and inexpensive. It is efficient and robust for any type of car parking zone and can be implemented without affecting its operation routine. The added slot booking, and navigation capabilities give user the flexibility and option to integrate the algorithm with other autonomous devices. The system proposed can save time and energy needed to search for vacant slots while providing with data to be integrated with autonomous vehicles.

## II. IMAGE PRE-PROCESSING AND ENHANCEMENT

The basic definition of image processing refers to processing of digital image, i.e. removing the noise and any kind of irregularities present in the image.

In image preprocessing, image data recorded restrain errors related to geometry and brightness values of the pixels. These errors are amended utilizing appropriate numerical models which are either definite or statistical models. Image enhancement is the alteration of image by changing the pixel brightness levels to enhance its visual effect. One of the basic improvements utilized is the Gaussian Blur technique.

A Gaussian blur is the consequence of blurring an image by using a Gaussian function. It is a generally utilized effect in graphics software, commonly to reduce image noise and decrease sharpness. The visual effect of this blurring method is a smooth blur similar to that of viewing the image through a translucent screen.

### III. IMAGE THRESHOLDING

Image segmentation is the method of dividing a digital image into various segments. The objective of segmentation is to classify as well as change the representation of an image into something that is more significant and simpler to examine. Image segmentation is typically used to locate objects and boundaries.

Usually Canny thresholding is preferred for edge detection and thresholding images in to respective binary form. Canny edge detection is a multi-step algorithm that can identify edges with noise suppressed at the same time.

### IV. MORPHOLOGICAL OPERATIONS

Morphological image processing is a collection of non-linear operations identified with the shape or morphology of features in an image. Morphological tasks depend just on the relative ordering of pixel values, not on their numerical values, and in this manner are particularly suited to the processing of binary images.

Morphological techniques test an image with a small shape or layout called a structuring element. The structuring element is situated at all possible areas in the image and it is contrasted and the corresponding neighborhood of pixels. Operations test whether the component "fits" inside the area, while others test whether it crosses the neighborhood.

### V. DATA HANDLING

The importing and exporting of information is the autonomous or semi-autonomous input and output of information between different applications. It involves changing from the format used by one application into another, where such change is accomplished autonomously via machine processes, such as transcoding, data transformation, and others. To utilize information delivered by another application. The capacity to import information is important in software applications since it implies that one application can complement another.

### VI. METHODOLOGY

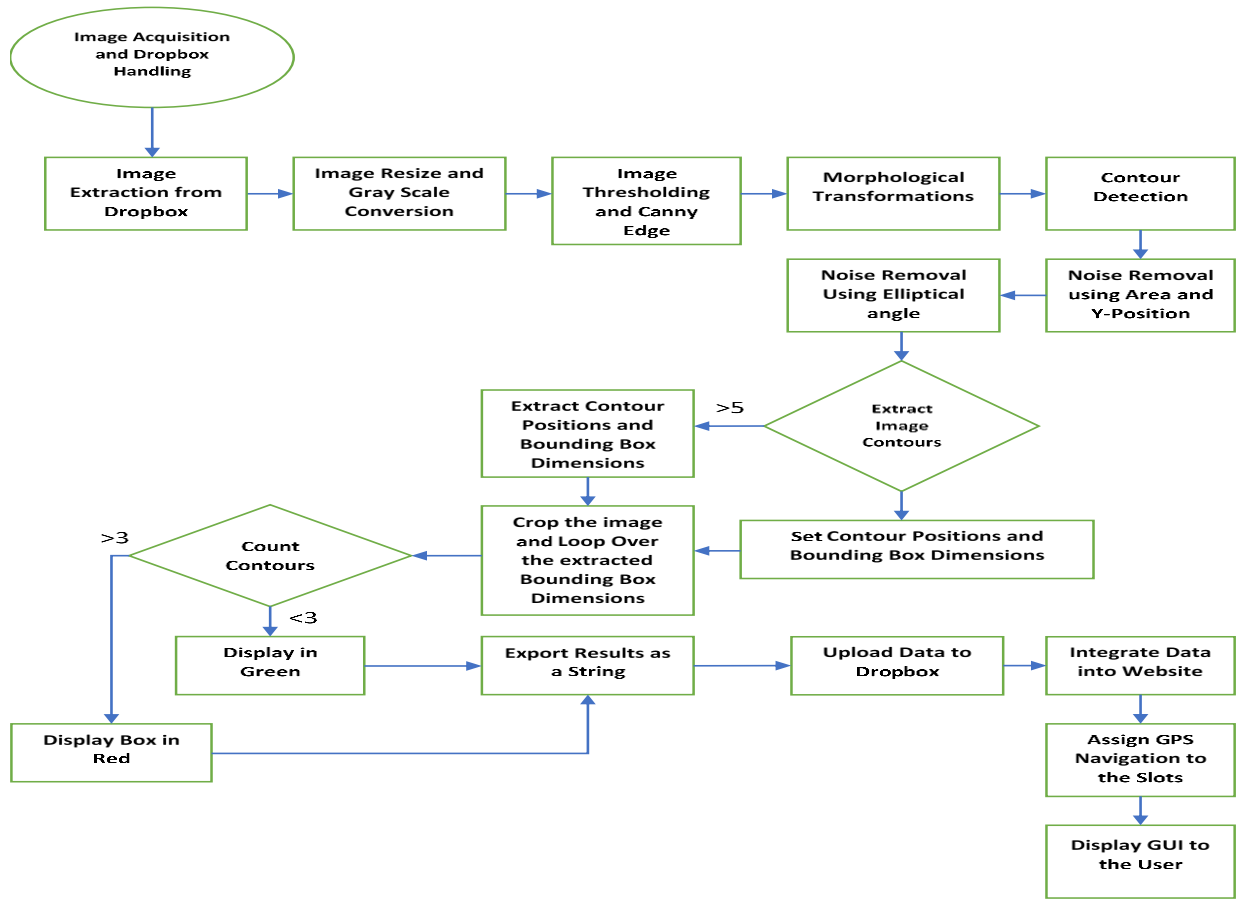

Figure 1: Methodology

Figure1 displays the methodology followed for the proposed method

## A. Image Acquisition and Conversion

Image acquisition is the first step in any visual based projects. The values and properties of the images acquired play an important role on how the data should be handled or dealt with.

In this paper, A practical and feasible acquisition method using the data storage capabilities of Dropbox is used to store and export images. Dropbox is cloud storage service that enables users to store files on remote cloud servers and the ability to share files within a synchronized format.

---

Algorithm 1

Output: Image saved in local disk

1. Generate Access Grant Token
2. Sync images from various Sources with Dropbox
3. Assign unique ID to images
4. Set Image Source and Destination Path in Raspberry Pi
5. Wait for user response
6. Extract images to Raspberry Pi using Dropbox Python Package
7. Check errors and store image in local disk
8. Update Dropbox data

---

Algorithm 1: Dropbox Image Handling and Extraction

Algorithm 1 displays the methodology used for storing and extracting images from dropbox to raspberrypi. The unique access grant key generated acts as an identifier to store the images and to access them. With the help of Dropbox package designed for use in python, one can seamlessly access and store data in Dropbox.

The image stored in the Dropbox are each assigned a unique id to differentiate them. The images are then stored in the specified path of the raspberrypi local memory to be used for further processing.

The images stored are then resized to the pixel value of width 960 pi and height 540 pi. The rescaling of images helps avoid any errors which can be generated due to the size disparity of various images when processed continuously. The images acquired from the process detailed in Algorithm 1 are the converted to their respective gray images which removes the dependency of the detection on the colors of the car. Removing the dependency of detection based on the colors of the car produces more accurate results when dealing with vehicles that have similar color to the backdrop of the image.

## B. Image Segmentation and Smoothing

Image acquired form the process detailed In Algorithm2 is then subjected to gaussian blurring using a 3x3 kernel to smooth the image and reduce high intensity value pixels. The Gaussian smoothing operator is a 2-D convolution operator that is used to `blur' images and remove detail and noise. It utilizes a disparate kernel that shows the shape of a Gaussian hump.

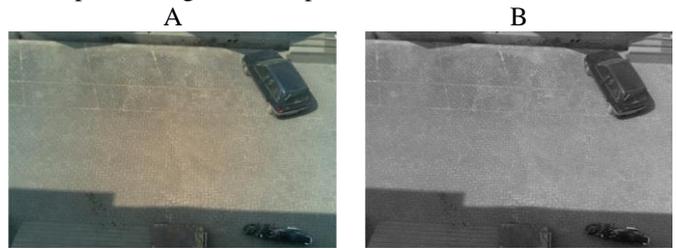

Figure 2A Gaussian approximation Function, Figure 2B Gaussian Blur Kernel

Figure 2A illustrates the Gaussian function or approximation with zero mean and σ defines the standard deviation of the approximation. Figure 2B defines the kernel used to perform the blurring operation on the images. The kernel is an 3x3 matrix which is convolved with the image to produce the desired blur effect. The standard deviation and mean values for the convolution are calculated in relation to the matrix while performing the blur operations.

Figure 3A input Image, Figure 3B Image after Gaussian Blur

It can be observed from Figure 3B that the image is now subjected to smoothing which removes the high intensity components of the image and gives a uniform distribution.

## C. Image Thresholding and Morphological Operations

Segmentation segregates an image into definite regions including each pixel with same features. For image investigation and interpretation, the locales should firmly relate to show objects or features of interest. The segmentation is carried out using truncate Thresholding of image in Figure 3B. In this type of thresholding, the destination pixel(fd(x,y)) is set to the threshold if the source pixel(fs(x,y)) value is greater than the threshold. Otherwise it is set to the source pixel value, maximum value is ignored.

$$fd(x,y) = \begin{cases} threshold, & fs(x,y) > threshold \\ fs(x,y), & Otherwise \end{cases}$$

Equation 1: Truncate Thresholding Principle

Equation 1 depicts the working of the segmentation procedure adopted for the algorithm. The image obtained after thresholding is subjected to canny Thresholding for additional extraction of the image features.

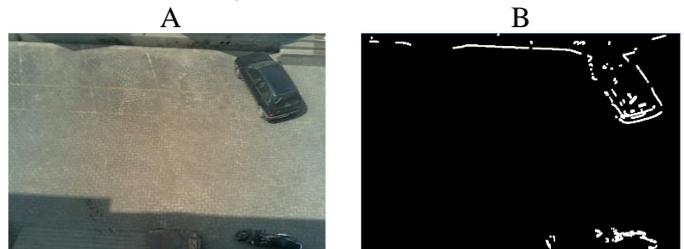

Figure 4A: input Image, Figure 4B: Binary Image

Figure 4b shows the image after truncate and canny thresholding. It can be observed that the Figure 4B shows the contours of the with some additional noise.

*D. Morphological Transformation and Contour Detection*

Morphological transformations such as erosion and dilation are performed on figure 4B to enhance the existing contour features. The erosion of a figure 4B by a structuring element produces a new binary image with ones in all locations (a,b) of a structuring element's origin at which that structuring element fits the Figure 4B. While dilation of the eroded image by a structuring element produces a new binary image with ones in all locations (a,b) of a structuring element's origin at which that structuring element hits the input image, repeating for all pixel coordinates (a,b). Dilation has the opposite effect to erosion as it adds a layer of pixels. Erosion is iterated once while Dilation is iterated twice with kernel of size 1x2 consisting of only ones.

Morphological transformation is used to enhance the features of the cars in the Figure 4B Binary Image. Two linear structuring elements and a diamond shaped structural element are created, which are an essential part of morphological dilation and erosion operations. A flat structuring element is a binary valued neighborhood, in which the true pixels are included in the morphological computation while excluding the false pixels. The center pixel of the structuring element, identifies the pixel in the image being processed.

$$k = \begin{bmatrix} 1 & 1 \end{bmatrix}$$

Equation 2: Kernel Value

'k' in Equation 2 represents the kernel used for the dilation and erosion processes. The kernel of size 2x1 was selected since sized sufficiently to fulfill the morphological operations without disturbing the neighboring contours.

The image after the enhancement is used to detect the contours. The contours of an image are connected components which are either 8 connected or 4 connected pixels. The retrieval method used, RETR_EXTERNAL, extracts the extreme outer contours present and the contour approximation algorithm, CHAIN_APPOX_NONE stores absolutely all the contour points.

*E. False Contour Removal and Module Classification*

It can be observed that while the use of gaussian blurring and canny thresholding can reduce the amount of noise in the Binary image, there is still a considerable amount of noise in the image. In this paper, an efficient algorithm for removal of the residual noise is proposed. The noise removal or false contour removals is based on the contour properties such as area and angle of the contours.

The first step includes classifying contours based on the area and Y-axis positions. Contours with area less than 70 units and Y-positions greater than 270 are made zero. The image is then classified based on the elliptical angles. Contours with angles between 80 and 100 are assumed to be noise and made zero.

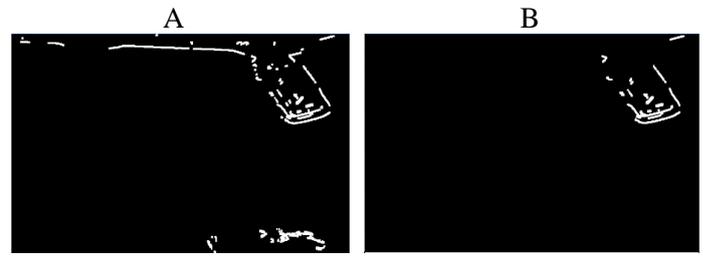

Figure 5A: Image before noise removal, Figure 5B: image after noise removal

It can be observed that there is a substantial decrease in the number of unwanted white pixels in the image. Figure 6B shows the contours related to the vehicle by eliminating other false contours.

The Algorithm is then split into two main modules depending on the number of contours. Module 1 is classified for working when the number of contours is five or greater. The algorithm is designed to extract the position of the vehicle contours from figure 5B and design the appropriate bounding boxes. The angle, height, and width are the extracted information from the contours, the number of iterations is also calculated depending on the extracted information for efficient processing.

Module 2 is designed to work when the number of contours detected is less. The bounding box properties and the loop iteration are manually assigned for the figure 5B. It has been made less autonomous since this module deals with the case of an empty parking slots. To avoid complications and errors during the detection of empty parking slots, manual inputs are given.

*F. Flexible Position Cropping*

The feature of the contours extracted from figure 5B are processed to determine the accurate cropping dimensions for the figure. Even after the noise from the image is removed, there are still a lot of empty spaces or null values in the image which are of no use. By removing the excess space, the computing can be made faster.

The contour with the highest Y-axis position is determined and used as an anchor to crop the image. Before designating the cropping ratio, the y value is equated to a threshold for error free handling. The position cropping is taken place only if the value of the cropping element is less than the limit. This process is only used with module 1, since there are no contours detected in module 2.

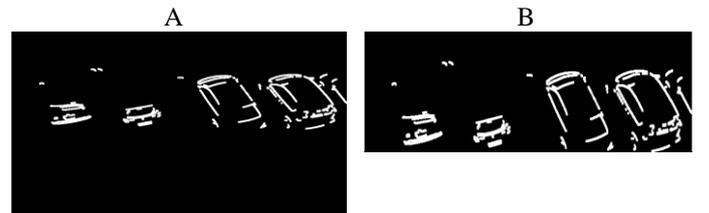

Figure 6A Image before Cropping, Figure 6B Image After Cropping

It can be observed that in Figure 6B the number of vacant spaces is reduced. This reduces the computational process and increases the efficiency.

*G. Data and Result Exporting*

The results obtained from the algorithm are uploaded to dropbox with the help of the access grant key. The data uploaded to dropbox is in form of image and text file. The output image consists the slots marked as empty or occupied, the text file in correspondence to the image file displays a string of 0's and 1's. The zeros correspond to empty slots while the ones correspond to occupied slots.

| Algorithm 2 |
|---|
| Input: Output files from Raspberry PI<br>Output: Display and Navigation for Vacant Slots<br><br>1. Acquire results<br>2. Extract the detected slot data into string and write the results to disk<br>3. Export results to Dropbox<br>4. Integrate data into website using PHP commands<br>5. Split the data in the string into individual slot data<br>6. Design slot interface and GUI using the extracted data<br>7. Assign navigation and reserve capabilities for empty slots<br>8. Display website GUI with user interfacing capability |

Algorithm 2: Uploading Algorithm

Algorithm 2 details the proposed method for exporting results to dropbox and to integrate it with the website. The website displays the results in an interactive GUI where the users can choose to reserve the empty slots and get directions to the slots.

## VII. EXPERIMENTAL RESULTS

The Figure 6B is iterated over using the information extracted for the values of iterations and Bounding Boxes. The Figure 6B is iterated in a continuous loop until the bounding boxes reach the image edge. The bounding box while iterating check for the number of contours present inside the box. The spot is declared vacant if the number of contour inside the bounding area is less than a certain threshold, else it is stated as occupied. The results produced are accurate since the bounding boxes are angled according to the cars. This inclination allows the bounding element to accurately assess the contours in its area.

Assessing accurate bounding boxes id crucial for the detection algorithm to work accurately. The modules are divided such that the values with the least error are used for constructing the bounding boxes. Module 1 is further split into two cases for accurate data extraction from the contours present. Cases 1 deals with when there's a car present in the first slot. This information is used as base for further iterations. Case 2 is considered when there is no car present in the first slot or no contours are detected during the first iteration. During case 2 manual values are assumed until the iteration or bounding boxes detects a car. Which then extracts the information from the detected contours to work with.

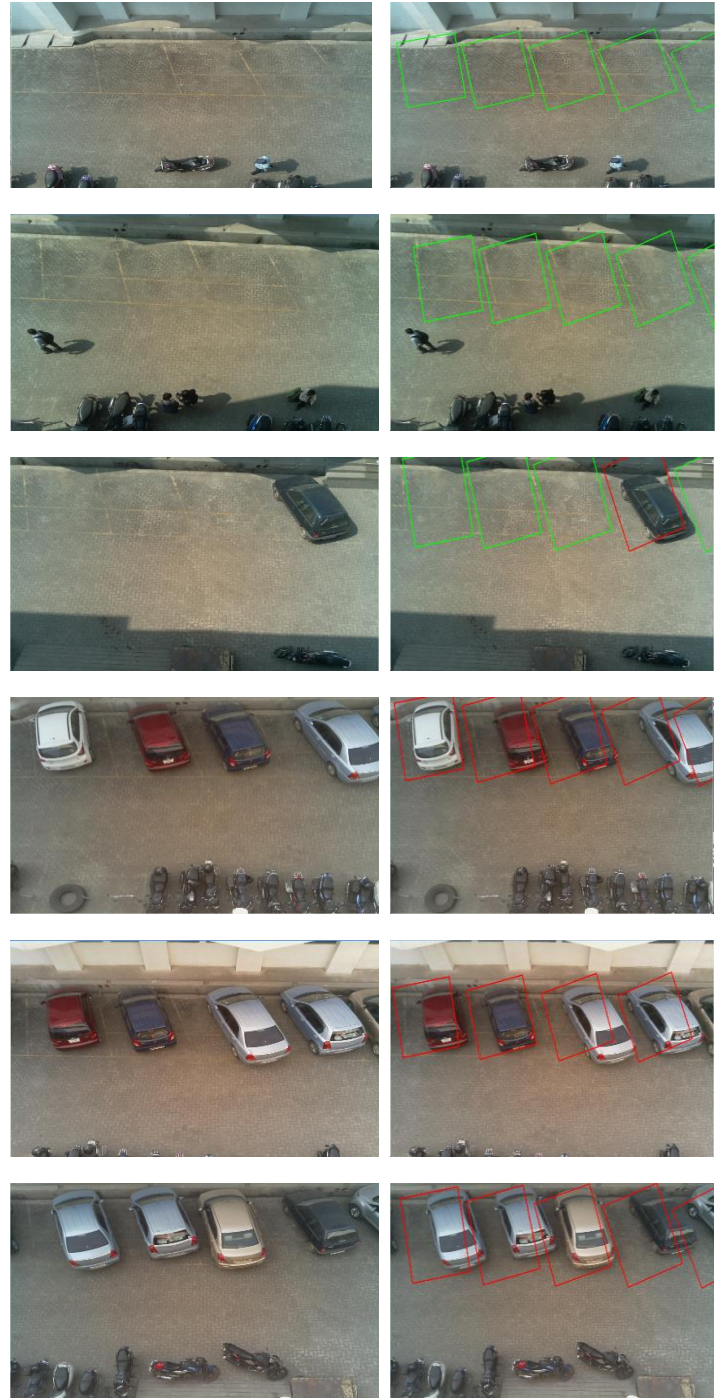

Input image(a)     Output Results(b)

Figure 7a shows the input image used, Figure 7b shows the output obtained

From Figure 7b we can observe that the results are shown as red and green bounding boxes. Bounding box is shown as red if a vehicle is present else it is displayed as green. It can also be observed that even though the cars are mot placed in their respective slots, the proposed algorithm detects them without any false results. The involvement of other objects or humans does not affect the results of the proposed algorithm.

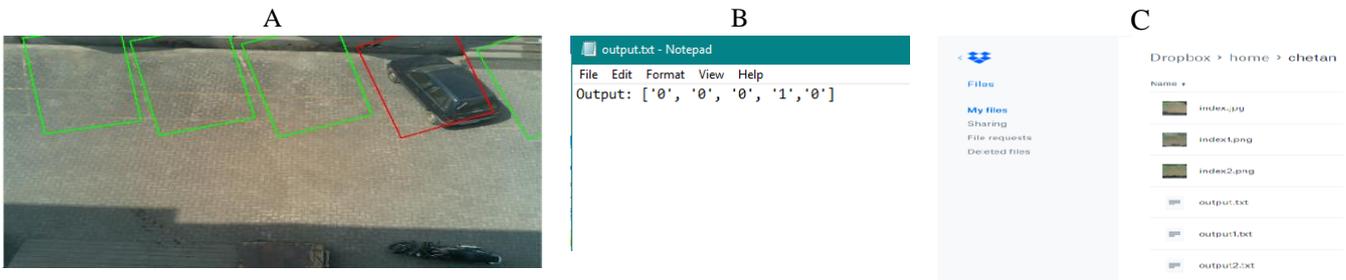

Figure 8A displays the output, Figure 8B shows the text file extracted, Figure 8C Data as stored in Dropbox

It can be observed in Figure 8B that the text file has 4 slots information while omitting the fifth detected slot which is a false detection. This correction step is implemented to maintain the size of the string for feasible transfer of data from dropbox(figure 9C) to the website server. In Figure 8C 'index' are the image results and 'output' are the text files with slot data.

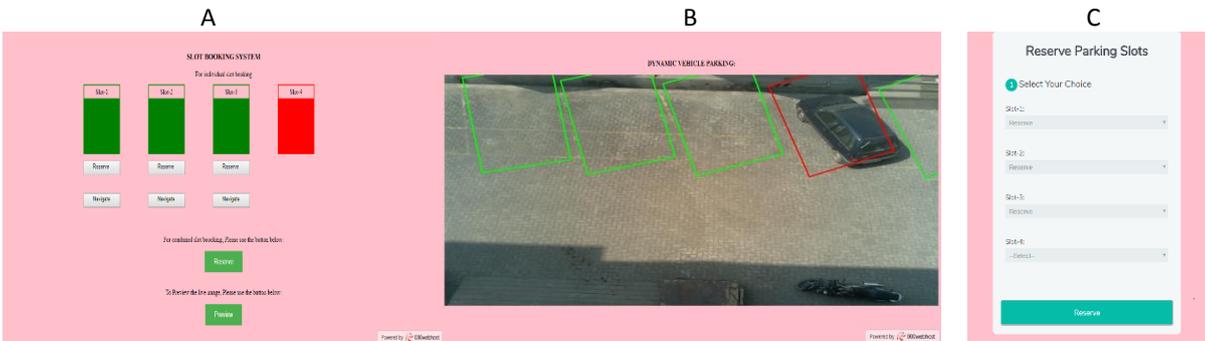

Figure 9A: website GUI, Figure 9B: results as displayed in website, 9C: Combined Slot Booking

Figure 9A displays the website GUI used to display the results to the user. It uses a slot like system to convert the string of data to be visualized by the user. The user is presented with options to reserve the slot one at a time or altogether as displayed in Figure 9C. The user is also give the option to get the destination GPS points and navigation options to the destination.

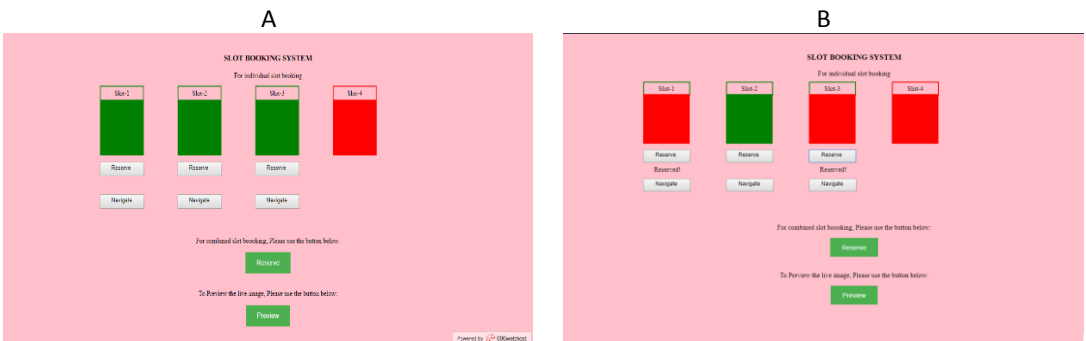

Figure 10 A: Slots before reservation, Figure 10B: Slots after reservation

It can be observed that in figure 10B that the slots which were reserved are displayed in red with a text stating they have been reserved. The reserve capability is a unique feature added to the project to facilitate easy and hassle-free parking. User can book the required free slots before arriving to the site and reduce time needed for searching a vacant spot.

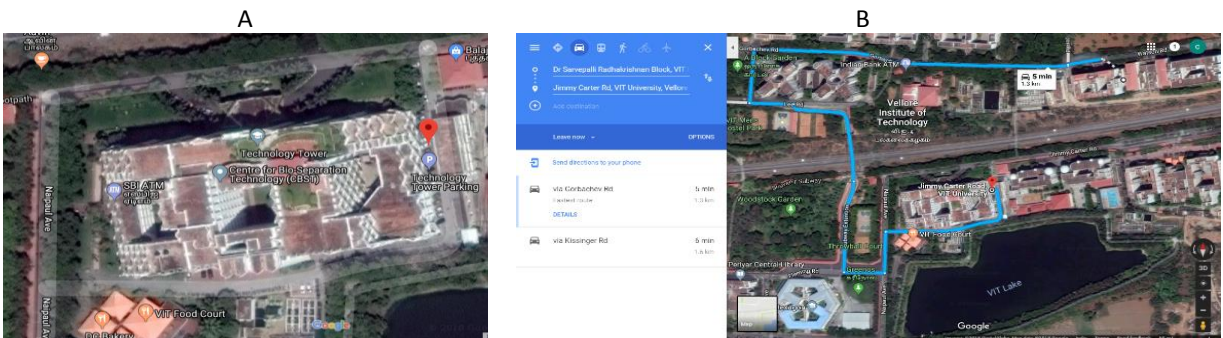

Figure 11A: Location map of the vacant spots, Figure 11B Route mapped towards the vacant spot.

Figure 11A displays the GPS location points assigned to the vacant parking slots. The user has the option to view the location of the vacant slots, which can be fed to autonomous vehicle or other devices for destination identification. Figure 11B displays the route to the destination selected. Instead of manually searching for slots, the proposed algorithm has the ability to find the vacant slots and route towards it, with the help of google maps.

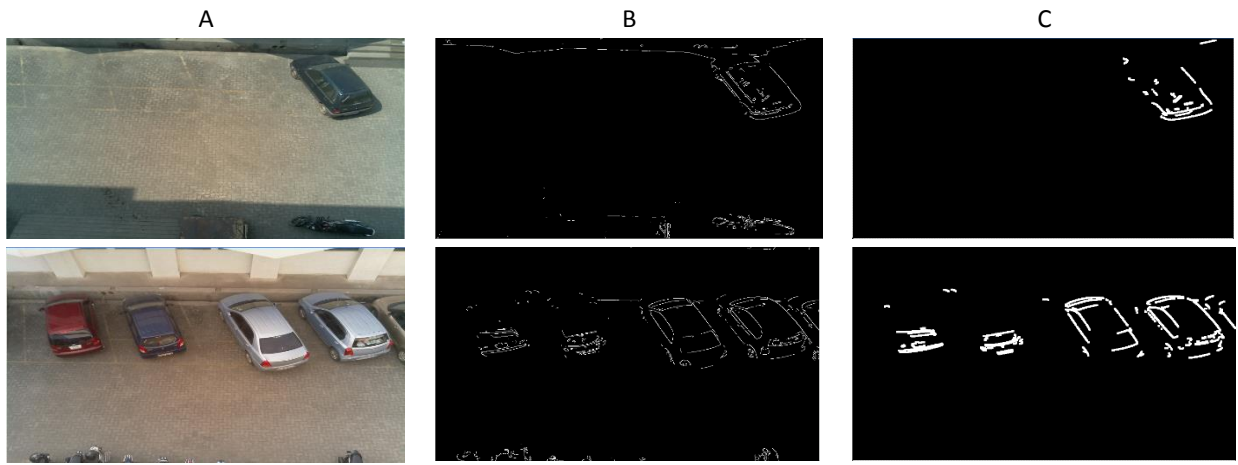

Figure 12A: Input Image, Figure 12B: Using canny alone, Figure 12 C: Proposed Algorithm

It can be observed that using the proposed algorithm 12C the contours of the vehicle are more prominent, and less noise is observed than in Figure 12B canny method. Using the proposed algorithm more accurate results are obtained due to less noise and enhancement of prominent features.

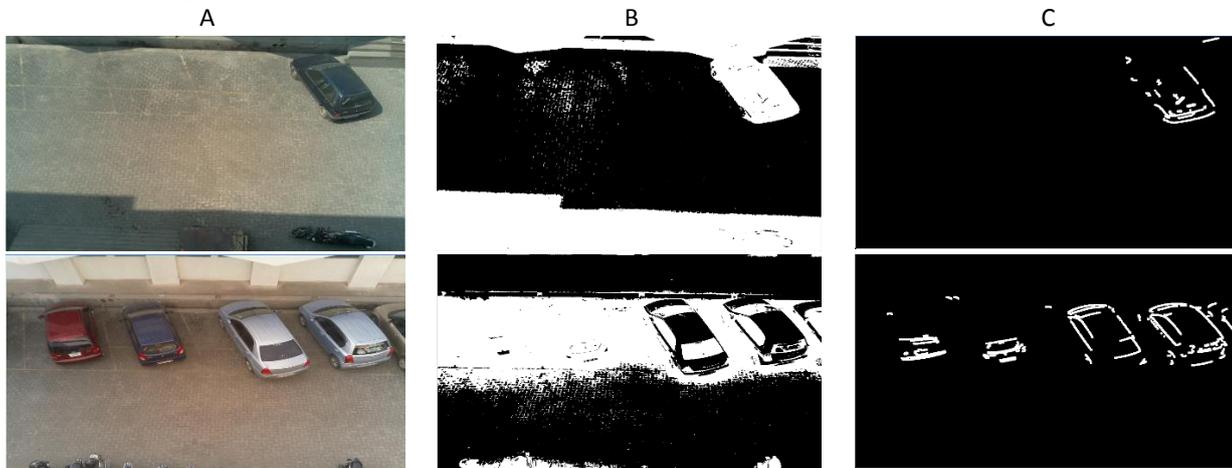

Figure 13A: Input Image, Figure 13B: Using inverse Binary, Figure 13 C: Proposed Algorithm

It can be seen in Figure 13b using inverse binary segmentation that most of the vehicles are not detected. The vehicles detected are merged into the background making it difficult to segregate. Using the proposed method Figure 13C we get clear cut contours of the image with complete detections of the vehicle.

To test the performance of our proposed algorithm, the efficiency of the system is measured with images taken at different time intervals. The performance is calculated by comparing the results of occupancy to the ground truth value. The performance of the proposed system is measured by the using the equation 3. Four Slots per test is considered for the experiment.

Number of Total Slots=4*(Tests Performed)
Number of Correct Slot Detections=4*(Correct Detections in the Tests Performed)
Accuracy percentage = (Number of correct slots detected/ Total slots) *100
Equation 3: Accuracy Percentage Calculations

Table 1

| Vehicle appearance | Tests Performed | Correct Detections | False Detection | Accuracy |
|---|---|---|---|---|
| Clear | 109 | 109 | 0 | 100% |
| Occluded | 139 | 138 | 1 | 99.28% |

From Table 1: The accuracy of the proposed algorithm is found to be 100%, 99.28% accurate during clear sky and occluded settings. Excess noise merged with the contours of the car and low contour detection, decreases the efficiency and the accuracy for detections during occluded settings. It is observed that the average performance is 99.64 % and is very high as compared with other parking lot detections applications. The accuracy of the proposed work also depends on the type of camera used for monitoring the parking lot.

Table 2

|  | Block Based Classification [12] | Edge Detection [13] | Twin ROI [14] | Inverse Binary Segmentation | Proposed method |
| --- | --- | --- | --- | --- | --- |
| Detection % | 97% | 98% | 95% | 60% | 99.64% |

From Table 2 we can say, the accuracy of our proposed system is superior than the Block Based Classification[12], Edge Detection[13], Twin ROI[14], and Inverse Binary Segmentation methods used in existing parking and image segmentation techniques. In other methods, the efficiency goes down when the car and parking area is of the same color and when less contours are detected.

## VIII. CONCLUSION

Parking slot detections with conventional algorithms give false positives in areas where the slot lines are distorted and when the vehicles overlap with slots itself. Even the amount of excess noise i.e. random vehicles and people loitering lead to false detection. The algorithm proposed was designed to combat these situations. The algorithm combines the image processing capabilities of OpenCV with the flexibility of Raspberry Pi for feasible and efficient Detection algorithms. Using various modules and cases made the algorithm robust and enables it to handle real life cases. The algorithm proposed has an efficient false contour detection and elimination technique to remove false predictions.

The algorithm proposed also includes data exporting and importing capabilities from selected servers. This allows the data to flow from dynamic points and to be integrated in to the algorithm. The results are displayed on a website where the user has the option to book slots or to set navigation path to the desired vacant spot.

This project can be integrated with self-driving and other autonomous vehicles for time efficient parking of vehicles. With the proposed method the user can select the desired spot for parking and give the data to any autonomous mobile device via website or the app. This allows for saving time and prevents the hassle of searching for parking spaces in large and crowded areas.

While the results produced are accurate, testing with existing autonomous is yet to be done. The manual setting during empty parking spaces can be replaced with dynamic algorithm with support for much larger ground. The angle and height of the image can be adjusted for more space availability and slot detections. The segmentation can be further improved for detection of complete vehicle contours.